\documentclass{amsart}

\usepackage{algorithmic}
\usepackage{listings}
\usepackage[norelsize,ruled]{algorithm2e}
\usepackage{tikz}
\usepackage{caption}
\usetikzlibrary{arrows}
\usepackage{amssymb, amscd, amsfonts}
\usepackage[mathscr]{eucal}
\usepackage{mathrsfs}
\usepackage{graphicx}
\usepackage{subfigure}
\graphicspath{{img/}}
\usepackage{lscape}
\usepackage{booktabs}

\newtheorem{theorem}{Theorem}[section]

\theoremstyle{definition}
\newtheorem{definition}[theorem]{Definition}

\newtheorem{claim}{Claim}

\theoremstyle{remark}

\newcommand{\remove}[1]{}

\newcommand{\Z}{{\mathbb{Z}}}

\begin{document}

\title[Eulerian digraphs and Dyck words, a bijection]{Eulerian digraphs and Dyck words, a bijection}


\author{Pietro Codara, Ottavio M. D'Antona, Marco Genuzio}


\date{\today}


\address[P. Codara, O. M. D'Antona, M. Genuzio]
{Dipartimento di Informatica, Universit\`{a} degli Studi di Milano
via Comelico 39/41, I-20135 Milano, Italy}
\email[P. Codara, O. M. D'Antona]{\{codara,dantona\}@di.unimi.it}
\email[M. Genuzio]{marco.genuzio@studenti.unimi.it}

\begin{abstract}
The main goal of this work is to establish a bijection between Dyck words and
a family of Eulerian digraphs. We do so by providing two
algorithms implementing such bijection in both directions. The connection
between Dyck words and Eulerian digraphs exploits a novel combinatorial
structure: a binary matrix, we call \emph{Dyck matrix}, representing the cycles
of an Eulerian digraph.
\end{abstract}
\maketitle

\section{Background, and motivation}
\label{sec:intro}
A digraph $G$ is \emph{Eulerian} if at every vertex the in-degree equals
the out-degree. (Note that we do not require $G$ to be connected.)
The edge set of an Eulerian digraph $G$ can be partitioned into directed cycles.
For a non-empty multiset $\mathbf{s}=\{s_1,s_2,\dots,s_n\}$, of $n$ positive integers,
we call an Eulerian digraph \emph{$\mathbf{s}$-labelled} if
its edge set is partitioned into $n$ directed cycles of length $s_1$, $s_2$, \dots, $s_n$, each with a distinguished first edge (and hence a unique second, third, etc., $m$-th edge).
Figure \ref{fig:E321} shows a $\{3,2,1\}$-labelled Eulerian digraph, with its 3 directed cycles of size
$1$, $2$ and $3$; the $j^{th}$ edge of the $i^{th}$ cycle is labelled $e_{i,j}$.
(Notice that in next Sections we endow these digraphs with a linear order on their cycles.)

\begin{center}
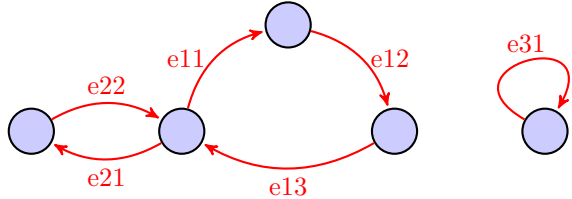

\begin{tikzpicture}[->,>=stealth',shorten >=1pt,auto,node distance=2cm,
thick,main node/.style={circle,fill=blue!20,draw,font=\sffamily\Large\bfseries}]
\node[main node,minimum size=.6cm] (1) {};
\node[main node,minimum size=.6cm] (2) [right of=1]{};
\node[main node,minimum size=.6cm] (3) [above right of=2]{};
\node[main node,minimum size=.6cm] (4) [below right of=3]{};
\node[main node,minimum size=.6cm] (5) [right of=4]{};
\path[color= red, every node/.style={color=red}]
(1) edge [bend left] node[above] {e22} (2)
(2) edge [bend left] node[below] {e21} (1)
(2) edge [bend left] node[left] {e11} (3)
(3) edge [bend left] node[right] {e12} (4)
(4) edge [bend left] node[below] {e13} (2)
(5) edge [in=60,out=150,loop]node[above] {e31} (5);
\end{tikzpicture}
\captionof{figure}{A $\{3,2,1\}$-Eulerian digraph.}
\label{fig:E321}
\end{center}

\smallskip
A \emph{Dyck word} on the alphabet\footnote{The choise of the alphabet follows \cite{galvin}.} $\{x,D\}$
is a string with the same number of $x$'s and
$D$'s, and such that the number of $x$'s in any initial segment is greater or equal to
the number of $D$'s. A \emph{Dyck path} is a lattice path in $\Z^2$ starting at $(0,0)$, ending on the
diagonal $y=x$, with unit steps in the North and East directions, and such that it does not pass below
the $y=x$ diagonal. One can easily build a correspondence between Dyck paths and Dyck words, by mapping a
North step to the character $x$ and an East step to the character $D$. Figure \ref{fig:dyck} shows the
Dyck path associated with the word $xxxDDxDDxD$. We denote by $\mathscr{W}$ the set of all Dyck words
on the alphabet $\{x,D\}$.

\begin{center}
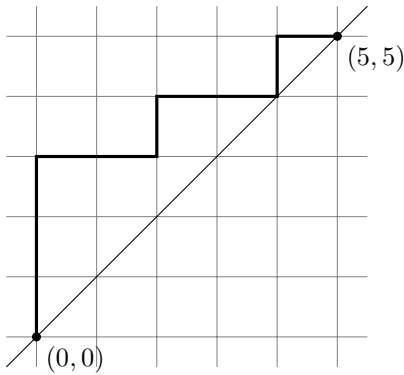

\begin{tikzpicture}
  \draw[step=0.8, help lines] (-0.4,-0.4) grid (4.4,4.4);
  \draw[very thick] (0,0)--(0,0.8)--(0,1.6)--(0,2.4)--(0.8,2.4)--
    (1.6,2.4)--(1.6,3.2)--(2.4,3.2)--(3.2,3.2)--(3.2,4)--(4,4);
  \draw[below right] (0,0) node(O){$(0,0)$};
  \draw[below right] (4,4) node(E){$(5,5)$};
  \fill (0,0) circle (0.06cm) (4,4) circle (0.06cm);

  \draw[-]
    (-0.4,-0.4)--(4.4,4.4);
\end{tikzpicture}
\captionof{figure}{The Dyck path $xxxDDxDDxD$.}
\label{fig:dyck}
\end{center}

\smallskip
The main goal of this work is to establish a bijection between a family of
$\mathbf{s}$-labelled Eulerian digraphs and the set $\mathscr{W}$.
To this end, we provide two algorithms implementing such bijection in both directions.
The connection between Dyck words and Eulerian digraphs exploits a novel combinatorial structure: a binary
matrix, we call Dyck matrix, representing the cycles of an Eulerian digraph.

\smallskip
Think of the symbol $x$ as multiplication times $x$, and of the symbol $D$ as
differentiation with respect to $x$. For each Dyck word $w$, the sequence of Stirling
numbers of the second kind of $w$, written $S_w(k)$, can be defined as the (unique) sequence
satisfying the following identity
\begin{equation}\label{eq:stirlings}
w f(x)=\sum_{k\geq 0}S_w(k) x^k D^k f(x)\,.
\end{equation}
Observe that, for $w=(xD)^n$, the unique sequence satisfying (\ref{eq:stirlings}) is the
sequence $S(n,k)$ of Stirling numbers of the second kind. Thus, the sequences $S_w(k)$
generalize Stirling numbers of the second kind. For more details
please refer to \cite{flajolet,galvin}, and references therein.

The results obtained in this work are the basis for a description of
Stirling numbers of Dyck words in terms of Eulerian digraphs.

\smallskip
The $\mathbf{s}$-labelled Eulerian digraphs are investigated in \cite{pavol}, in the
special case of $n$ cycles of length $m$.
There, the authors  show that $\mathbf{s}$-labelled Eulerian digraphs with $k$ vertices
represent a combinatorial interpretation of the generalized Stirling numbers
$S_{m}(n,k)$ introduced in \cite{blasiak1}. Specifically, in \cite[Theorem 3.1]{pavol} they prove
the correspondence between $\mathbf{s}$-labelled Eulerian digraphs and colourings (hence stable
partitions) of
disjoint unions of cliques of the same size.

\smallskip
Later, Engbers, Galvin, and Hilyard (\cite{galvin}) pointed out that the structures
investigated in \cite{pavol} constitute a combinatorial interpretation of the
Stirling numbers of the second kind associated with Dyck words of the form $w=(x^m D^m)^n$.

\section{A family of Eulerian digraphs, and their cycle matrices}
\label{sec:bijection}

The main data structure our algorithms are based on is a binary matrix,
defined as follows.


\begin{definition}\label{def:matrix}
A non-empty binary matrix $M=(m_{i,j})$ of size $n\times k$, is a \emph{Dyck matrix} if it satisfies the
following conditions.
\begin{itemize}
\item[(M1)] There exists $0<h\leq k$ such that $m_{1,j}=1$ if and only if $j\leq h$.
\item[(M2)] For each $1 \leq i < n$, there exist $1\leq a_i\leq b_i<c_i\leq k$ such that $a_i$ is
    the smallest index satisfying $m_{i,a_i}=1$ and $m_{i+1,a_i}=0$, $b_i$ is the greatest index
    such that $m_{i,b_i}=1$, and
    $c_i$ is the greatest index such that $m_{i+1,c_i}=1$. Moreover, the following hold: \\
    (M2.1) $m_{i+1,j}=m_{i,j}\,$,\ \ for $j=1,\dots,a_i -1$;\\
    (M2.2) $m_{i+1,j}=0\,$,\ \ for $j=a_i,\dots,b_i$;\\
    (M2.3) $m_{i+1,j}=1\,$,\ \ for $j=b_i+1,\dots,c_i$;\\
    (M2.4) $m_{i+1,j}=0\,$,\ \ for $j=c_i+1,\dots,k$.
\end{itemize}
\end{definition}

\smallskip
For example, the matrix
\begin{equation}\label{eq:matrix}
\begin{pmatrix}
  1 & 1 & 1 & 0 & 0\\
  1 & 0 & 0 & 1 & 0\\
  0 & 0 & 0 & 0 & 1
\end{pmatrix}
\end{equation}
is a Dyck matrix, while the matrix
\[
\begin{pmatrix}
  1 & 1 & 1 & 0 & 0\\
  1 & 0 & 0 & 1 & 1\\
  0 & 0 & 0 & 0 & 1
\end{pmatrix}
\]
is not, in that it violates condition (M2).

\smallskip
A Dyck matrix $M=(m_{i,j})$ of size $n\times k$ can be associated to an Eulerian digraph $E$ with
$n$ cycles and $k$ vertices, endowed with an order on its cycles. For $i=1,\dots,n$, the $i^{th}$ row
represents the $i^{th}$ cycle $C_i$. Specifically, if $U=\{j_1,j_2,\dots,j_{s_i}\}$ is the non-empty set
of indices such that $m_{i,j}=1$ if and only if $j\in U$, then $v_{j_1},v_{j_2},\dots,v_{j_{s_i}}$ are the vertices
of $C_i$, and $(v_{j_1},v_{j_2}),(v_{j_2},v_{j_3}),\dots,(v_{j_{s_i-1}},v_{j_{s_i}}),(v_{j_{s_i}},v_{j_1})$
are its edges. If $U$ is a
singleton, then $C_i$ is a loop. We say that $M$ is the \emph{cycle matrix} of $E$. One can easily check that
the matrix represented in (\ref{eq:matrix}) is the cycle matrix of the Eulerian digraph depicted in
Figure \ref{fig:E321}.

We can characterize the family $\mathscr{E}_W$ of Eulerian digraph associated to Dyck matrices, as follows.
Let $\mathbf{s}=(s_1,s_2,\dots,s_n)$, with $s_i>0$ for each $i=1,\dots,n$.
Denote by $C_1, \dots, C_n$ the cycles of $E$. The $\mathbf{s}$-labelled Eulerian digraph $E$
belongs to $\mathscr{E}_W$ if and only if the following conditions hold.
\begin{itemize}
\item[(E1)] No cycle is contained into another cycle.
\item[(E2)] If two cycles $C_i$, $C_{i+1}$ share $k$ vertices, these must be the first $k$
vertices of both cycles.
\end{itemize}

\smallskip
Finally, we are ready to introduce our main results. Denote by ${\mathscr M}_W$ the class of all Dyck matrices.
\begin{theorem}
\label{th:main}
$\mathscr{W}$ and ${\mathscr M}_W$ are in bijection.
\end{theorem}

The proof of Theorem \ref{th:main} is provided in Section \ref{sec:proof}. Next we introduce
the algorithms to associate Dyck words with cycle matrices of ordered Eulerian digraphs, hence
Dyck matrices, and viceversa.

\section{From Dyck words to Eulerian digraphs}
\label{sec:dyck_to_eulerian}

We supply an online algorithm that converts a Dyck word in the corresponding Dyck matrix.
The algorithm implements an incremental construction of the result.
Here, the idea is to split a Dyck word into \emph{slopes}, {\it i.e.} maximal continuous
sequences of $x$'s, and \emph{descents}, {\it i.e.} maximal continuous sequences of $D$'s (see \cite{galvin}).
We call \emph{valley} a descent followed by a slope, and \emph{peak} a slope followed by a descent.
Every peak represents a cycle.
For every cycle the number of $x$'s from the beginning of the $slope$ to its end is
the number of new vertices, with respect to the previous cycle, while the difference between the number
of $D$'s and the number of $x$'s represent the number of nodes shared with the next cycle.
Our algorithm incrementally builds the matrix, generating a new row at the end of each descent.

\smallskip
The function \texttt{getMatrix}, implementing the online conversion algorithm is shown below.
The function receives in input a stream of characters, the Dyck word,
and outputs the associated Dyck matrix.

In line $1$ we initialize the four variables $\mathcal{X}$, $\mathcal{D}$, $k$, and $prev$.
$\mathcal{X}$ and $\mathcal{D}$ are counters for the number of
$x$'s and $D$'s, respectively; the variable $k$  represents the number of nodes shared by
two consecutive cycles; $prev$ is the last character read.
In lines $2$--$10$ we process the stream of characters until its end (\texttt{EOS}), checking that
$\mathcal{D}$ never exceeds $\mathcal{X}$ (if $\mathcal{X} < \mathcal{D}$ the string does not
represent a Dyck word). Line $3$ reads the next character of the stream: if this is a $D$, we set $prev$ to $``D"$ and increase $\mathcal{D}$.
Otherwise (that is, if the next character is $x$), we check (line $5$) whether the $x$ follows
an $x$ or a $D$. In the former case, we simply increase $\mathcal{X}$, while in the
latter we can create a new $row$ and insert it in the Dyck matrix.
The creation of a new row (line $7$, and line $11$) implements the following steps.
\begin{itemize}
\item[(R1)] We create a copy $R$ of the previous row. We modify $R$ maintaining only the first
$k$ $1$'s, and resetting (to $0$) the others. (If we are creating the first row, $R$ is an empty array.)
\item[(R2)] We append to $R$ a sequence of $\mathcal{X}-k$ $1$'s.
\end{itemize}
The new $row$ will have $\mathcal{X}$ $1$'s.
In line $8$ we insert the new $row$ into the matrix $M$. We also need to append a sequence of $\mathcal{X}-k$
$0$'s to each previous row.
In line $9$, we update all the variables, setting $prec$ to ``x'', $k$ to $\mathcal{X}-\mathcal{D}$,
$\mathcal{X}$ to $k+1$, and $\mathcal{D}$ to $0$. The last row is created and added to the matrix in lines $11$--$12$.

\smallskip
\IncMargin{1em}
\begin{algorithm}[H]
\label{alg:1}
\Indm
 \SetKwFunction{Conv}{getMatrix}
\Conv{s}{
 \Indp
\SetKwFor{While}{do}{}{while}
\SetKwFor{While}{do}{}{while}

\nl  $\mathcal{X}=0$\,; $\mathcal{D}=0$\,; $k=0$\,; $prev=``x"$\;
\nl  \Repeat{NextChar\textup{(}$s$\textup{)} $\neq$ \textup{\texttt{EOS}}
        \ \textup{\texttt{and}}\ \ $\mathcal{X} \geq \mathcal{D}$ }{
\nl  \eIf{NextChar($s$) is $``D"$}{
\nl   $prev=``D"$;  $\mathcal{D}=\mathcal{D}+1$\;
}{
\nl \eIf{ $prev=``x"$}{
\nl $\mathcal{X}=\mathcal{X}+1$\;
}{
\nl create a new $row$ \;
\nl add $row$ to $M$ and fill previous rows with $0$'s\;
\nl $prec=``x"$\,; $k=\mathcal{X}-\mathcal{D}$\,; $\mathcal{X}=k+1$\,;  $\mathcal{D}=0$\;
}}
\nl remove first character of $s$ \;
}
\nl create last $row$\;
\nl add last $row$ to $M$ and fill previous rows with $0$'s\;
\nl \Return $M$\;
}
\end{algorithm}
\DecMargin{1em}

\section{From Eulerian digraphs to Dyck words}
\label{sec:eulerian_to_dyck}

Here, we supply an algorithm to convert a Dyck matrix in a Dyck word. The algorithm implements an incremental
construction of the Dyck word. The main idea is to scan the whole matrix using a
vertical two-value window (a $2\times 1$ binary matrix), to check the values of two elements lying in
consecutive rows, but in the same column.
For practical reasons, we add a ``virtual'' row of $0$'s as a first row and as a last row of the matrix.
After this operation we can start scanning the matrix, moving horizontally our windows till the end
of each row, then skipping to the next row. Whenever the windows identifies a combination\footnote{For
convenience, we denote our two-value matrix by its transposed.} $(0,1)$, we can append an
$x$ to the Dyck word we are building, because such pair represents a vertex that does not belong
to any previous cycle.
The pair $(0,0)$ does not give any information. The pair $(1,1)$ represents a node belonging to two consecutive
cycles. The pair $(1,0)$ denotes a vertex that belongs to a cycle, but does not to the next cycle:
whenever our windows identifies this combination we append a $D$ to the Dyck word.

\medskip
The function \texttt{getDyckWord} receives in input a Dyck matrix $M=(m_{i,j})$. (The algorithm
can also deal with a binary stream, representing the matrix read row by row.)

In line $1$ we initialize the two variables $i$ (the counter for rows) and $dyck$ (the Dyck word to output).
(Here, $\epsilon$ denote the empty string.)
Then, after adding (line $2$) two rows of $0$'s in the first and last position of $M$,
we scan the matrix (line $3$--$8$) checking the values of the windows $(m_{i-1,j},m_{i,j})$.
The algorithm uses the function \texttt{checkWindow}$(a,b)$, which takes two binary values and returns:
\begin{itemize}
\item $``D"$, if $a=1$ and $b=0$;
\item $``x"$, if $a=0$ and $b=1$;
\item $\epsilon$, otherwise.
\end{itemize}

In the code below, \texttt{EOR} denotes the End Of a Row, while \texttt{EOM} denotes the End Of the Matrix.
The symbol $\ast$ is used to denote the catenation of two strings.
\IncMargin{1em}
\begin{algorithm}[h]
\label{alg2}

\SetAlgoLined
\SetKwFunction{Conv}{getDyckWord}
\SetKwFunction{Couple}{checkWindow}
\SetKwProg{myalg}{Algoritmo}{}{}
\Indm
\Conv{$M=(m_{i,j})$}
\Indp\\
\nl $i=1$; $dyck=\epsilon$\;
\nl Add a row of $0$'s as a first and last row of $M$\;
\nl \Repeat{\textup{\texttt{EOM}}}{
\nl $j=0$\;
\nl \Repeat {\textup{\texttt{EOR}}}{
\nl $dyck=dyck$ $\ast$ \Couple$(m_{i-1,j},m_{i,j})$\;
\nl $j=j+1$\;
}
\nl $i=i+1$\;
}
\nl \Return $dyck$\;
%
%
%
\end{algorithm}
\DecMargin{1em}

\section{Proof of Theorem \ref{th:main}, and further remarks.}
\label{sec:proof}

To prove Theorem \ref{th:main} we show that the algorithm described in Sections \ref{sec:dyck_to_eulerian}
and \ref{sec:eulerian_to_dyck} are corrects, in that they associate a Dyck word with a Dyck matrix, and
viceversa. Moreover, we show that both algorithms implement injective maps, and that one map is the inverse of the other.

\bigskip
\textbf{From Dyck words to Dyck matrices.} The algorithm writes a row of the matrix at each valley, that is, when an
$x$ is read after a $D$. When the first valley is met (or at the end of the word, if there are no valleys) a first
row is written, with $\mathcal{X}$ $1$'s (the number of $x$ read, from the beginning of the word). Such row
satisfies condition (M1) in Definition \ref{def:matrix}, and it will do so also when, in line
$8$, we append to the row a number of sequences of $0$'s.
For every subsequent valley encountered, the algorithm writes a row with the following properties.

\smallskip
(A) The row has $h$ $1$'s below the ones of the previous row, and $h$ is smaller than the total number of $1$'s
of the previous row. The positions of such $1$'s coincide with the positions of the first $h$ $1$'s of the
previous row (lines $8$ and $11$, described in (R1), Section \ref{sec:dyck_to_eulerian}).

\smallskip
(B) The row has a sequence of $u>1$ adjacent $1$'s, starting at the index $b+1$, where $b$
is the position of the last $1$ in the previous row
(lines $8$ and $11$, described in (R2), Section \ref{sec:dyck_to_eulerian}).

\smallskip
(C) The row has a number of $0$'s at the end, since the algorithm appends a number of sequences of $0$'s (line $8$).

\smallskip
From (A), (B), and (C), we easily derive that the matrix satisfies the property (M2) in Definition \ref{def:matrix}.
By construction, the mapping of Dyck words into Dyck matrices is injective.

\begin{figure}[h]
\centering
\begin{tabular}{ccccc}
$xxxDD$ & \quad\quad & ${\color{gray}xxxDD}xDD$ & \quad\quad & ${\color{gray}xxxDDxDD}xD$ \\
\quad & \quad & \quad & \quad & \\
$\begin{pmatrix}
1&1&1
\end{pmatrix}$ & \quad\quad &
$\begin{pmatrix}
{\color{gray}1}&{\color{gray}1}&{\color{gray}1}&0\\
1&0&0&1
\end{pmatrix}$ & \quad\quad &
$\begin{pmatrix}
{\color{gray}1}&{\color{gray}1}&{\color{gray}1}&{\color{gray}0}&0\\
{\color{gray}1}&{\color{gray}0}&{\color{gray}0}&{\color{gray}1}&0\\
0&0&0&0&1
\end{pmatrix}$
\end{tabular}
\caption{From $xxxDDxDDxD$ to its cycle matrix, step by step.}\label{fig:algosteps}
\end{figure}

\bigskip
\textbf{From Dyck matrices to Dyck words.}

Let $M=(m_{i,j})$ be a Dyck matrix of size $n\times k$, and let $M'$ be the matrix
obtained from $M$ by adding a first row and
a last row of $0$'s, according with line $2$ of the algorithm in Section \ref{sec:eulerian_to_dyck}.

\begin{claim}\label{claim}
Every column of $M'$ is formed by
\begin{itemize}
\item[(i)] an initial non-empty segment of $0$'s, followed by
\item[(ii)]a non-empty sequence of $1$'s, followed by
\item[(iii)]a non-empty sequence of $0$'s.
\end{itemize}
\end{claim}
To prove Claim \ref{claim}, first observe that (i) trivially follows by the fact that
we have added an initial row of $0$'s. From condition (M2.3)
in Definition \ref{def:matrix} we deduce that every column must contain at least a $1$. Condition (M2.1)
in the same Definition say that every $1$ can be followed by other $1$'s, and every $0$ by other $0$'s.
Summarizing, we have that every column has an initial segment formed by a non-empty sequence of $0$'s
followed by a non-empty sequence of $1$'s. The addition of a final row of $0$'s
implies that each column has a final non-empty sequence of $0$'s.

It remains to show that if in a column a $1$ is followed by a $0$,
then no other $1$'s appear in the subsequent positions of the column.
Let $m_{r,s}$ be the first $0$ that follows a $1$ in the $s^{th}$ column of $M$, if such element exists (if not,
our claim about $M'$ is trivially verified for the column $s$). Consider again Definition \ref{def:matrix}:
by (M2.3), together with the fact that $b_{r-1}<c_{r-1}$ (M2), there exists $j>s$ such that $m_{r,j}=1$. Hence,
for each $i>r$, the condition in (M2.3) does not apply for the elements $m_{i,s}$. Instead, only conditions in (M2.1) or (M2.2) can apply: in both cases $m_{i,s}=0$.

\bigskip
The algorithm in Section \ref{sec:eulerian_to_dyck} scans $M'$ by rows using a vertical $2\times 1$ window.
Whenever the algorithm hits the combination $(0,1)$, it appends an $x$ to a word $W$ (starting by the empty word),
and whenever it hits the combination $(1,0)$, it appends a $D$. By Claim \ref{claim},
$W$ starts with $x$, and contains exactly $n$ $x$'s and $n$ $D$'s.
Moreover, at any position the number of $D$'s can never exceed the number of $x$'s.
Indeed, the scan can not hit the combination $(1,0)$ before hitting $(0,1)$ on the same column.
Hence, $W$ is a Dyck word.

\bigskip
We do not prove here that the map implemented
by the algorithm is injective.
We show, instead, that the maps implemented by the two algorithms are one the inverse of the other.

Let $W$ be a Dyck word with at least one valley (if there are no valley, our claim easily follows). We follow row by row the action of the function \texttt{getMatrix} on input $W$, and the action of
\texttt{getDyckWord} on the rows that are written. The aim is to show that \texttt{getDyckWord} returns exactly the word $W$.

Recall that \texttt{getMatrix} writes a row of the matrix at each valley. When
the first valley is met, a sequence of $1$'s is written in the first row (after that, a number of $0$'s will follow, till the end of the row). The number of these $1$'s equals the number $t$ of $x$'s of the first slope of $W$. Since the reverse algorithm adds a
row of $0$'s as a first row, when it scans the first two rows of its matrix it begins writing a Dyck word by appending $t$ $x$'s to the empty string.

When \texttt{getMatrix} finds a second valley (or the end of $W$), it writes the second row of the matrix, according to (A), (B), and
(C) in the previous paragraph. According to (A), the beginning of the row is a copy of the previous row. When scanning this part of the row,
the reverse algorithm will do nothing.

According to (A) and (B), the following part of the new row is formed by a sequence of $0$'s, and at least
one of this $0$'s lies below a $1$. In correspondence of such $0$'s the reverse algorithm will append some $D$'s to its output.
By (R1) in Section \ref{sec:dyck_to_eulerian} the number of $0$'s lying below $1$'s is $\mathcal{X}-k = \mathcal{D}$.
Thus, it coincides with the number of $D$ of the first descent of $W$.

According to (B), a non-empty sequence of $1$'s is written in the following part of the row.
Such $1$'s lie below $0$'s, and, by (R2), the number of such $1$'s equals the
number of $x$ in the second slope. Hence, when the reverse algorithm scans this part of the row, it appends to its output
string the correct number of $x$.

The same occurs until the last row of the matrix is written. At this point the output of the reverse algorithm
coincides with W without its final descent. But \texttt{getDyckWord} adds a final row of $0$'s to the matrix. When this final row is scanned,
\texttt{getDyckWord} appends a sequence of $D$'s to its output. Such $D$'s make the output string a correct Dyck word. Hence, the output
must coincide with $W$.

\section{Conclusion, and future work}
\label{sec:conclusion}
We have provided a bijection between the set of all Dyck words and a class of binary matrices, we call Dyck matrices. Dyck matrices are
the cycle matrices of a family of $\mathbf{s}$-labelled Eulerian digraphs, endowed with an order
on their cycles.

Further work aims to describe the Stirling numbers of a Dyck word (see \cite{galvin}) in terms of Eulerian digraphs.
Indeed, it seems possible, following the same approach as in \cite{pavol}, to obtain, for any Dyck word $w$,
the Stirling number $S_w$ as the collection of $\mathbf{s}$-labelled Eulerian digraph obtained by taking the
graph $E$ associated with $w$ by the algorithm described in Section \ref{sec:dyck_to_eulerian} and applying
appropriate transformations on $E$.

\bibliographystyle{amsalpha}
\bibliography{cdg_gascom}

\providecommand{\bysame}{\leavevmode\hbox to3em{\hrulefill}\thinspace}
\providecommand{\MR}{\relax\ifhmode\unskip\space\fi MR }
\providecommand{\MRhref}[2]{%
  \href{http://www.ams.org/mathscinet-getitem?mr=#1}{#2}
}
\providecommand{\href}[2]{#2}
\begin{thebibliography}{CDH14}

\bibitem[BF12]{flajolet}
P.~Blasiak and P.~Flajolet, \emph{Combinatorial models of
  creation-annihilation}, S\'eminaire Lotharingien de Combinatoire \textbf{65}
  (2010/12), Art. B65c, 78.

\bibitem[BPS03]{blasiak1}
P.~Blasiak, K.~A. Penson, and A.~I. Solomon, \emph{{The boson normal ordering
  problem and generalized Bell numbers}}, Annals of Combinatorics \textbf{7}
  (2003), no.~2, 127--139.

\bibitem[CDH14]{pavol}
P.~Codara, O.~M. D'Antona, and P.~Hell, \emph{{A simple combinatorial
  interpretation of certain generalized Bell and Stirling numbers}}, Discrete
  Mathematics \textbf{318} (2014), no.~1, 53--57.

\bibitem[EGH13]{galvin}
J.~Engbers, D.~Galvin, and J.~Hilyard, \emph{{Combinatorially interpreting
  generalized Stirling numbers}}, arXiv:1308.2666v3 [math.CO] (2013).

\end{thebibliography}

\end{document}